\documentclass{article}
\usepackage{frascatiphys,here,graphicx,subfigure}
\begin{document}
\title{ANISOTROPY IN THE COSMIC RADIATION AT TEV ENERGY
}
\author{
Roberto Iuppa\\
{\em University of Tor Vergata and INFN, sez.ne Roma Tor Vergata}\\
{\em via della Ricerca Scientifica 1, Rome - 00133, Italy}
}
\maketitle
\baselineskip=11.6pt
\begin{abstract}
In recent years very important results were obtained from cosmic ray
experiments about the arrival direction distribution of primaries in the
TeV energy range. As most of these particles are charged nuclei, they
are deflected by the magnetic field they pass through before reaching
the Earth surface, the effect of the Lorentz force being inversely
proportional to the particle energy. As far as the local interstellar
medium is known, the gyroradius of a 10 TeV proton is expected to be
only 100 a.u., small enough to make the arrival direction distribution
isotropic. Since 1930s a "large scale" (90°-120°) anisotropy  is known
to exist, generally interpreted as the combined effect of sources far
away and magnetic fields nearby. Nonetheless, in the last decade
experiments like Tibet-ASg, Milagro, ARGO-YBJ and IceCube discovered
structures as wide as 10°-30° all over the sky at ~10 TeV energy, what
is unexplainable within the standard model of cosmic rays.
In this paper a review of the most recent experimental results about
cosmic ray anisotropy is given, together with the status of the art of
theoretical efforts aimed at interpreting them within the current cosmic
ray paradigma.
\end{abstract}
\baselineskip=14pt
\section{Introduction}
As CRs are mostly charged nuclei, their paths are deflected and highly isotropized by the action of galactic magnetic field (GMF) they propagate through before reaching the Earth atmosphere. The GMF is the superposition of regular field lines and chaotic contributions and the local total intensity is supposed to be $B=2\div 4\textrm{ $\mu$G}$ \cite{beck01}. In such a field, the gyro-radius of CRs is given by ${r}_{a.u.}\approx 100_{\textrm{\scriptsize{TV}}}$, where $r _{a.u.}$ is in astronomic units and R$_{\textrm{\scriptsize{TV}}}$ is the rigidity in TeraVolt. 
Clearly, there is very little chance of observing a point-like signal from any radiation source below $10^{17}{\rm eV}$, as they are known to be at least several hundreds parsecs away.

If it is true that magnetic fields are the most important ``isotropizing'' factor when they randomly vary on short distances, it is clear as much that some particular features of the magnetic field at the boundary of the solar system or farther might focus CRs along certain lines and the observed arrival direction distribution turns out to be consequently an-isotropic.

Different experiments observed an energy-dependent \emph{``large scale''} anisotropy with amplitude spanning 10$^{-4}$ to 10$^{-3}$, from tens GeV to hundreds TeV, suggesting the existence of two distinct broad regions, an excess named \emph{``tail-in''} (distributed around 40$^{\circ}$ to 90$^{\circ}$ in Right Ascension (R.A.) and a deficit named \emph{``loss cone''} (distributed around 150$^{\circ}$ to 240$^{\circ}$ in R.A.).

Moreover, in the last decade smaller excesses ($\sim30^{\circ}$ wide) were found to exist in the CR arrival direction distribution.

The origin of the galactic CR anisotropy is still unknown, but the study of its evolution over the energy spectrum has an important valence to understand the propagation mechanisms and the structure of the magnetic fields through which CRs have traveled.
\section{Experimental results}
\label{sec:datareview}

\begin{figure}[!htbp]
  \centering
  \includegraphics[width=\textwidth]{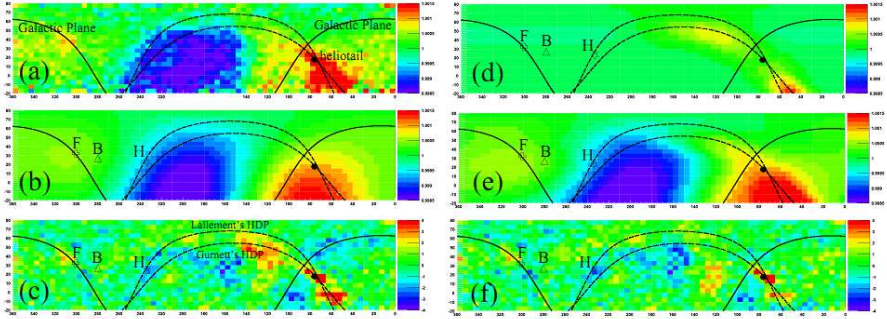}
  \caption{Anisotropy maps of galactic CRs observed and reproduced at the modal energy of 7 TeV by the Tibet-AS$\gamma$ experiment \cite{amenomori10}. (a): the observed CR intensity; (b): the best-fit large scale component; (c): the significance map of the residual anisotropy after subtracting the large scale component; (d): the best-fit medium scale component; (e): the best-fit large+medium scale components; (f): the significance map of the residual anisotropy after subtracting the large and the medium scale component. The solid black curves represent the galactic plane. The dashed black curves represent the Hydrogen Deflection Plane. The helio-tail direction is indicated by the black filled circle. The open cross and the inverted star with the attached characters ``F'' and ``H'' represent possible orientations of the local interstellar magnetic field. The open triangle with ``B'' indicates the orientation of the best-fit bi-directional cosmic-ray flow obtained in the reference \cite{amenomori10}.
    \label{fig:tibet_model}}
\end{figure}

In 2006 the Tibet AS$\gamma$ experiment, located at Yangbajing (4300 m a.s.l.), published the first 2D high-precision measurement of the CR anisotropy in the Northern hemisphere in the energy range from few to several hundred TeV \cite{amenomori06}. In the figure \ref{fig:tibet_model} the CR intensity map observed by Tibet AS$\gamma$ is shown (panel (a)), together with some theoretical model. The Tibet AS$\gamma$ collaboration carried out the first measurement of the energy and declination dependences of the R.A. profiles in the multi-TeV region with a single EAS array, revealing  finer details of the known anisotropy. They found that the first harmonic amplitude is remarkably energy-independent in the range 4 - 53 TeV and all the components of the anisotropy fade out for CR energy higher than a few hundred TeV, showing a co-rotation of galactic CRs with the local Galactic magnetic environment.

The Milagro collaboration published in 2009 a 2D display of the sidereal anisotropy projections in R.A. at a primary CR energy of about 6 TeV \cite{milagro09}. They observed a steady increase in the magnitude of the signal over seven years, in disagreement with the Tibet AS$\gamma$ results \cite{amenomori10}.
It is worth noting that the energy at which the Tibet AS$\gamma$ and Milagro results were obtained ($\sim$ 10 TeV) is too high for Sun effects play an important role.

In 2007, modeling the large scale anisotropy of 5 TeV CR, the Tibet-AS$\gamma$ collaboration ran into a ``skewed'' feature over-imposed to the broad structure of the so-called tail-in region \cite{amenomori07}. They modeled it with a couple of intensity excesses in the hydrogen deflection plane \cite{gurnett06,lallement05}, each of them 10$^{\circ}$-30$^{\circ}$ wide. A residual excess remained in coincidence with the helio-tail. See the figure \ref{fig:tibet_model} (d) and its caption for more details.
%
%

\begin{figure}[!htbp]
\centering
\includegraphics[width=\textwidth]{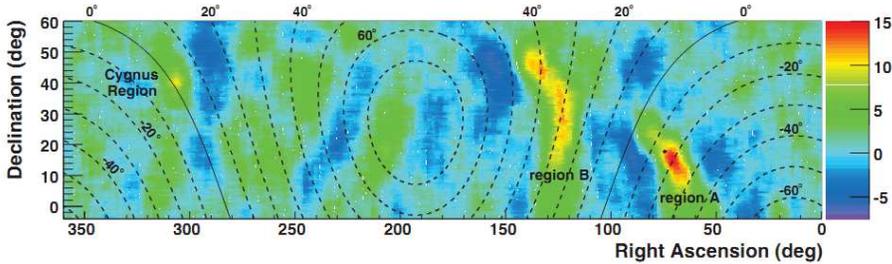}
    \caption{Significance map for the Milagro data set without any cuts to remove the hadronic CR background. A 10$^{\circ}$ bin was used to smooth the data, and the color scale gives the statistical significance. The solid line marks the Galactic plane, and every 10$^{\circ}$ in Galactic latitude are shown by the dashed lines. The black dot marks the direction of the helio-tail, which is the direction opposite the motion of the solar system with respect to the local interstellar matter. 
 \label{fig:milagro2008}}
\end{figure}

Afterwards the Milagro collaboration claimed the discovery of two localized regions of excess 10 TeV CRs on angular scales of 10$^{\circ}$ with greater than 12 $\sigma$ significance \cite{milagro2008}.
The figure \ref{fig:milagro2008} reports the pre-trial significance map of the observation. Regions ``A'' and ``B'', as they were named, are positionally consistent with the ``skewed feature'' observed by Tibet-AS$\gamma$.

The strongest and most localized of them (with an angular size of about 10$^{\circ}$) coincides with the direction of the helio-tail.
The fractional excess of region A is $\sim 6\times$10$^{-4}$, while for region B it is $\sim$ 4$\times$ 10$^{-4}$. The deep deficits bordering the excesses are due to a bias in the reference flux calculation. This effect slightly underestimates the significance of the detection.
The Milagro collaboration excluded the hypothesis of gamma-ray induced excesses.
In addition, they showed the excess over the large scale feature without any data handling (see the figure 2 of \cite{milagro2008}). 
%
%

The excesses in both regions are harder than the spectrum of the isotropic part of CRs. 

Easy to understand, more beamed the anisotropies and lower their energy, more difficult to fit the standard model of CRs and galactic magnetic field to experimental results. 
In addition, the observation of a possible small angular scale anisotropy region contained inside a larger one rely on the capability for suppressing the smooth global CR anisotropy at larger scales without, at the same time, introducing effects of the analysis on smaller scales.

Nonetheless, this observation has been confirmed by the ARGO-YBJ experiment\cite{argomedium2, argomedium3} at median energy of the isotropic CR proton flux of about E$_p^{50}\approx$1.8 TeV (mode energy $\approx$0.7 TeV)
\section{Models and interpretations}
\label{sec:models}
Some authors suggested that the large scale anisotropy can be explained within the diffusion approximation taking into account the role of the few most nearby and recent sources \cite{blasi11,erlykin06}.
Other studies suggest that a non-di-polar anisotropy could be due to a combined effect of the regular and turbulent GMF \cite{battaner09}, or to local uni- and bi-dimensional inflows \cite{amenomori10}. In particular the authors modeled the observed anisotropy by a superposition of a large, global anisotropy and a midscale one. The first one is proposed to be generated by galactic CRs interacting with the magnetic field in the local interstellar space surrounding the heliosphere (scale $\sim$2 pc).

About the medium scale anisotropy, no theory of CRs in the Galaxy exists yet which is able to explain few degrees anisotropies in the rigidity region 1-10 TV leaving the standard model of CRs and that of the local galactic magnetic field unchanged at the same time. All the solutions proposed to explain the phenomenon are too complex to fit in this work. For quite a complete review see \cite{anis-rev}.
\section{Conclusions}
Current experimental results show that the main features of the anisotropy are uniform in the energy range (10$^{11}$ - 10$^{14}$ eV). Structures are there in every region of the harmonic domain down to angular scales as narrow as $10^{\circ}$. 
So far, no theory of CRs in the Galaxy exists which is able to explain both large scale and few degrees anisotropies leaving the standard model of CRs and that of the local galactic magnetic field unchanged at the same time. 
\section{Acknowledgements}
The author wishes to thank Prof. R. Santonico and Dr. G. Di Sciascio for their support in data analysis and critical review of theoretical models.
\end{document}